\documentclass[aps,prl,twocolumn,superscriptaddress,showpacs]{revtex4}
\usepackage{graphicx}
\usepackage{dcolumn}
\usepackage{bm}
\usepackage{color}
\usepackage{float}
\usepackage{amsmath}
\usepackage{amssymb}

\newcommand{\eph}{{\it e-ph}}
\newcommand{\ee}{{\it e-e}}

\begin{document}
\title{Competing phases and orbital-selective behaviors in the two-orbital Hubbard-Holstein model}
\author{Shaozhi Li}
\affiliation{Department of Physics and Astronomy, The University of Tennessee, Knoxville, Tennessee 37996, USA}
\author{Ehsan Khatami}
\affiliation{Department of Physics and Astronomy, San Jos\'{e} State University, San Jos\'{e}, California 95192, USA}
\author{Steven Johnston}
\email{sjohn145@utk.edu}
\affiliation{Department of Physics and Astronomy, The University of Tennessee, Knoxville, Tennessee 37996, USA}
\affiliation{Leibniz-Institute for Solid State and Materials Research, Institute for Solid State Theory, IFW-Dresden, D-01171 Dresden, Germany}
\date{\today}

\begin{abstract}
We study the interplay between the electron-electron ({\ee}) and
the electron-phonon ({\eph}) interactions in the two-orbital Hubbard-Holstein model
at half filling using the dynamical mean field theory. We find that the {\eph}
interaction, even at weak couplings, strongly modifies the phase diagram of
this model and introduces an orbital-selective Peierls insulating phase (OSPI) that is
analogous to the widely studied orbital-selective Mott phase (OSMP). At small
{\ee} and {\eph} couplings, we find a competition between the OSMP and the
OSPI, while at large couplings, a competition occurs between Mott and
charge-density-wave (CDW) insulating phases. We further demonstrate that the Hund's
coupling influences the OSPI transition by lowering the energy associated with
the CDW. Our results explicitly show that one must be
cautious when neglecting the {\eph} interaction in multi-orbital systems, where
multiple electronic interactions create states that are readily influenced by perturbing interactions. 
\end{abstract}
\maketitle 

{\em Introduction} --- 
In recent years many researchers have focused on studying electron-electron
(\ee) interactions in multiorbital systems such as the iron-based
superconductors (FeSCs). In doing so, they have discovered numerous new
phenomena, including the Hund's metal
\cite{HauleNJP,Georges,Fanfarillo_PRB2015}  and the orbital-selective Mott
phase (OSMP) \cite{Anisimov,Koga,deMediciPRL2014,Yi_PRL2013, ShaozhiPRB2016},  
which arise from the competing action of the electronic interactions.
These concepts have helped shape our understanding of the enigmatic properties of these materials. Despite this success, however, surprisingly little is
currently known about how competition/cooperation with other factors such as
impurities or the electron-phonon (\eph) interaction influences these phenomena.
This question is important for our microscopic understanding of these
materials, as subtle multiorbital correlation effects can produce states that
are readily affected by small perturbations. 
 
In the case of the FeSCs, the {\eph} interaction was ruled out as a possible
pairing mediator by early {\em ab initio} calculations \cite{BoeriDFT}
indicating that the total coupling strength was small, with a dimensionless
{\eph} coupling $\lambda \le 0.2$. Because of this, many researchers have
assumed that this interaction plays a secondary role in these materials with
regards to other aspects as well. However, there is growing evidence that this
outlook may have been premature. For example, more recent calculations find
that taking into account the possible magnetism
\cite{Yndurain_PRB,Boeri_PRB,CohPRB2016,MandalPRB2014} or orbital fluctuations
\cite{Kotani_PRL2010, Saito_PRB2010} can increase the total {\eph} coupling
strength compared to the original estimates. This finding is consistent with
the general notion that electron correlations can enhance {\eph} interactions
\cite{CaponeReview}.  Moreover, the discovery of the FeSe films on oxide
substrates \cite{WangCPL} has implicated new possible lattice interactions,
either across the interface \cite{Lee_nature2014, Rademaker_njop} or within the
FeSe film \cite{CohNJP2015}.  Since bulk FeSe is believed to be in the OSMP
regime \cite{FESE1,FESE2}, these experiments naturally raise questions about
when and how {\eph} interactions can influence such multi-orbital phenomena. 

Hubbard-Holstein models are the simplest models capturing the interplay between
{\ee} and {\eph} interactions. The single-band variant has been extensively
studied, particularly at half-filling, where a direct competition occurs
between antiferromagnetic Mott insulating (MI) and charge-density-wave (CDW)
phases \cite{Bauer_epl,  Bauer_prb, Berger_PRB, Nowadnick_PRL, Murakami_PRB,
Werner_PRB, SangiovanniPRL2006, MacridinPRL2006, Nocera, KhatamiPRB2008}. In comparison, far fewer studies exist for multiband generalizations of the model
\cite{Yamada_japan, Kotani_PRL2010}. Motivated by this, we carried out a
dynamical mean field theory (DMFT) \cite{Georges_mod} study of a degenerate
two-orbital Hubbard-Holstein model with inequivalent bandwidths. Here, we focus
on the half-filled case and construct low-temperature phase diagrams in the
$\lambda$-$U$ and $\lambda$-$J$ planes, where $U$ and $J$ are the Hubbard and
Hund's interaction strengths, respectively. Similarly to the single-band case, we
observe a competition between CDW and MI phases when the {\eph} and {\ee}
interactions are large. When the interactions are weak to intermediate in
strength, however, we find additional phases displaying orbital-selective
behavior.  The first is the now well-studied OSMP driven by the electronic
interactions \cite{deMediciPRL2011}.  The second is a lattice-driven analog of
the OSMP, which we refer to as an orbital-selective Peierls insulator (OSPI).
The phase boundaries of the model are also significantly influenced by the
{\eph} interaction, even for relatively weak values of $\lambda < 0.3$.  This
result clearly demonstrates that one cannot rule out the influence of the
{\eph} in correlated multiorbital systems {\em a priori} based on DFT-based
estimates for the total coupling strength. 

{\em Methods} --- 
The Hamiltonian for the degenerate two-orbital Hubbard-Holstein model
\cite{Yamada_japan} is
$H=H_\mathrm{kin}+H_\mathrm{lat}+H_\mathrm{e-ph}+H_\mathrm{e-e}$, where
\begin{eqnarray}
H_\mathrm{kin}&=&-\sum_{\langle i,j \rangle,\gamma,\sigma}t^{\phantom\dagger}_{\gamma}c^{\dagger}_{i,\gamma,\sigma}c^{\phantom\dagger}_{j,\gamma,\sigma}-\mu\sum_{i,\gamma,\sigma}\hat{n}_{i,\gamma,\sigma}, \nonumber\\
H_\mathrm{e-ph}+H_\mathrm{lat}&=&g\sum_{i,\gamma,\sigma}\left(b^{\dagger}_i+b^{\phantom\dagger}_i\right)\left(\hat{n}_{i,\gamma,\sigma}-\frac{1}{2}\right)+\Omega\sum_{i}b^{\dagger}_ib^{\phantom\dagger}_i,\nonumber\\
H_\mathrm{e-e}&=&U\sum_{i,\gamma}\hat{n}_{i,\gamma,\uparrow}\hat{n}_{i,\gamma,\downarrow}+U^\prime\sum_{i,\gamma\ne \gamma^\prime}\hat{n}_{i,\gamma\uparrow}\hat{n}_{i,\gamma^\prime,\downarrow}\nonumber\\
&&+(U^\prime-J)\sum_{i,\gamma<\gamma^\prime,\sigma}\hat{n}_{i,\gamma,\sigma}\hat{n}_{i,\gamma^\prime,\sigma}\nonumber\\
&&+J\sum_{i,\gamma\ne \gamma^\prime} ( c_{i,\gamma,\uparrow}^{\dagger}c_{i,\gamma,\downarrow}^{\dagger}c_{i,\gamma^\prime,\downarrow}^{\phantom\dagger}c_{i,\gamma^\prime,\uparrow}^{\phantom\dagger}\nonumber\\
&&-c_{i,\gamma,\uparrow}^{\dagger}c_{i,\gamma,\downarrow}^{\phantom\dagger}c_{i,\gamma^\prime,\downarrow}^{\dagger}c_{i,\gamma^\prime,\uparrow}^{\phantom\dagger} ) \nonumber
\end{eqnarray}
\noindent Here, $\langle\cdots\rangle$ denotes a summation over nearest neighbors; $c^{\dagger}_{i,\gamma,\sigma}$ creates an electron with spin $\sigma$ in orbital $\gamma=1,2$ on site $i$; $b^{\dagger}_i$ creates a phonon on site $i$; $\hat{n}_{i,\gamma,\sigma}=c^\dagger_{i,\gamma,\sigma}c^{\phantom\dagger}_{i,\gamma,\sigma}$ is the particle number operator;  $t_\gamma$ is the nearest neighbor hopping integral for orbital $\gamma$; $U$ and $U^\prime$ are the intra- and inter-orbital Hubbard interactions, respectively. Throughout, we choose $U^\prime=U-2J$ due to rotational symmetry \cite{Georges, Castellani_PRL}. $J$ is the Hund's coupling, which is fixed to $J = U/5$ unless otherwise stated; $g$ is the {\eph} interaction strength; $\Omega$ is the phonon energy;  and $\mu$ is the chemical potential, which is adjusted to fix the average particle per site to $\langle \hat{n}\rangle=2$. 

\begin{figure}
\center\includegraphics[width=0.95\columnwidth]{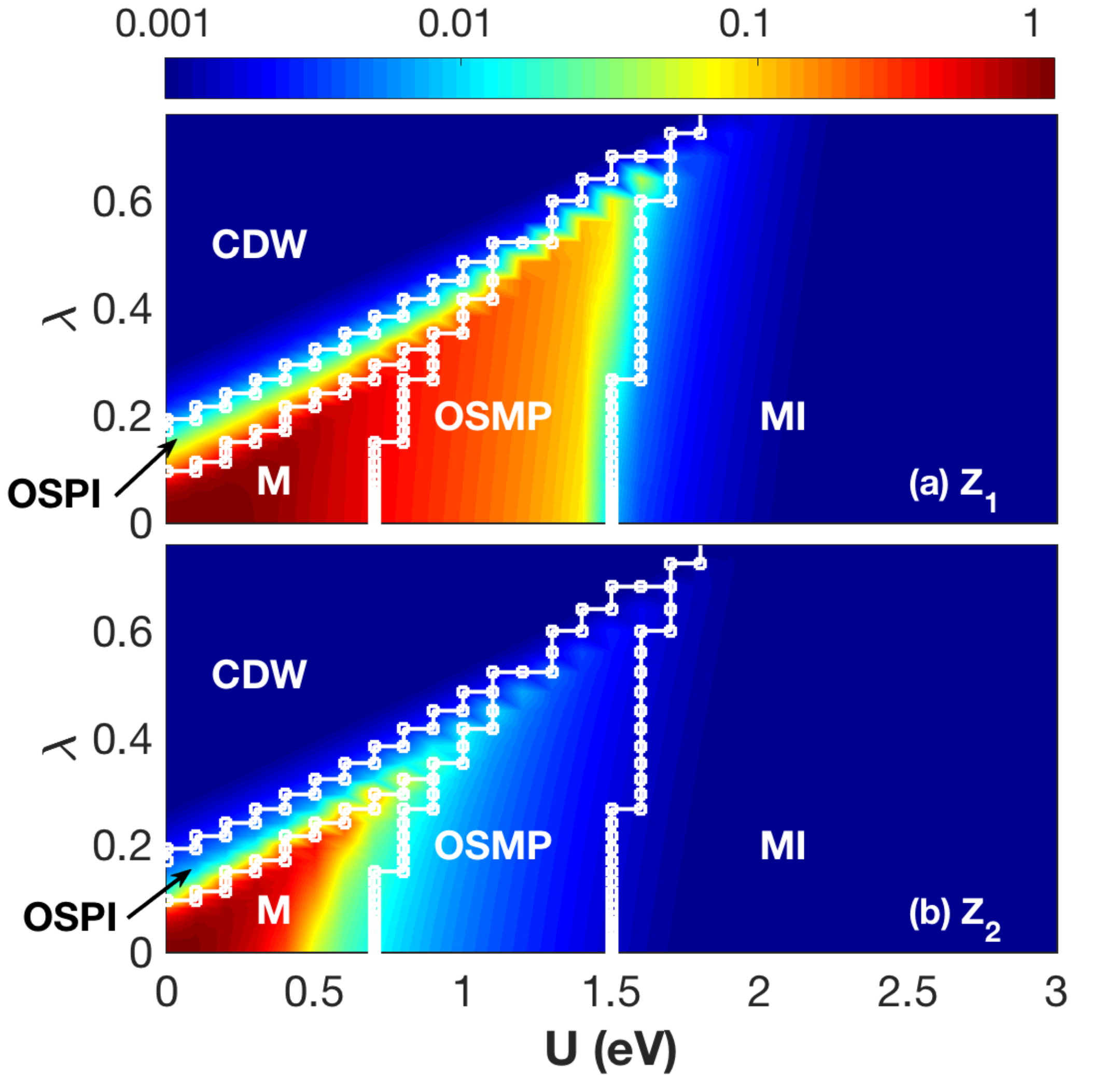}
\caption{\label{Fig:phasegu}{(color online)} The phase diagram for the two-orbital Hubbard-Holstein model in the {\eph} interaction strength ($\lambda$) - Hubbard $U$ plane at
 charge density $n=2$ and temperature $\beta=200/W$.
(a) and (b) show density plots of quasiparticle weights $Z_{1}$ and $Z_{2}$ on a logarithmic scale, respectively. The different phases are labeled as follows: metal (M),
orbital-selective Mott phase (OSMP), Mott insulater (MI), charge density wave (CDW), and orbital-selective Peierls insulator (OSPI). The white dots indicate points where the calculations were performed, and we plotted them to show phases boundaries. The color scale is plotted using a linear interpolation.}
\end{figure}

\begin{figure}
\center\includegraphics[width=0.75\columnwidth]{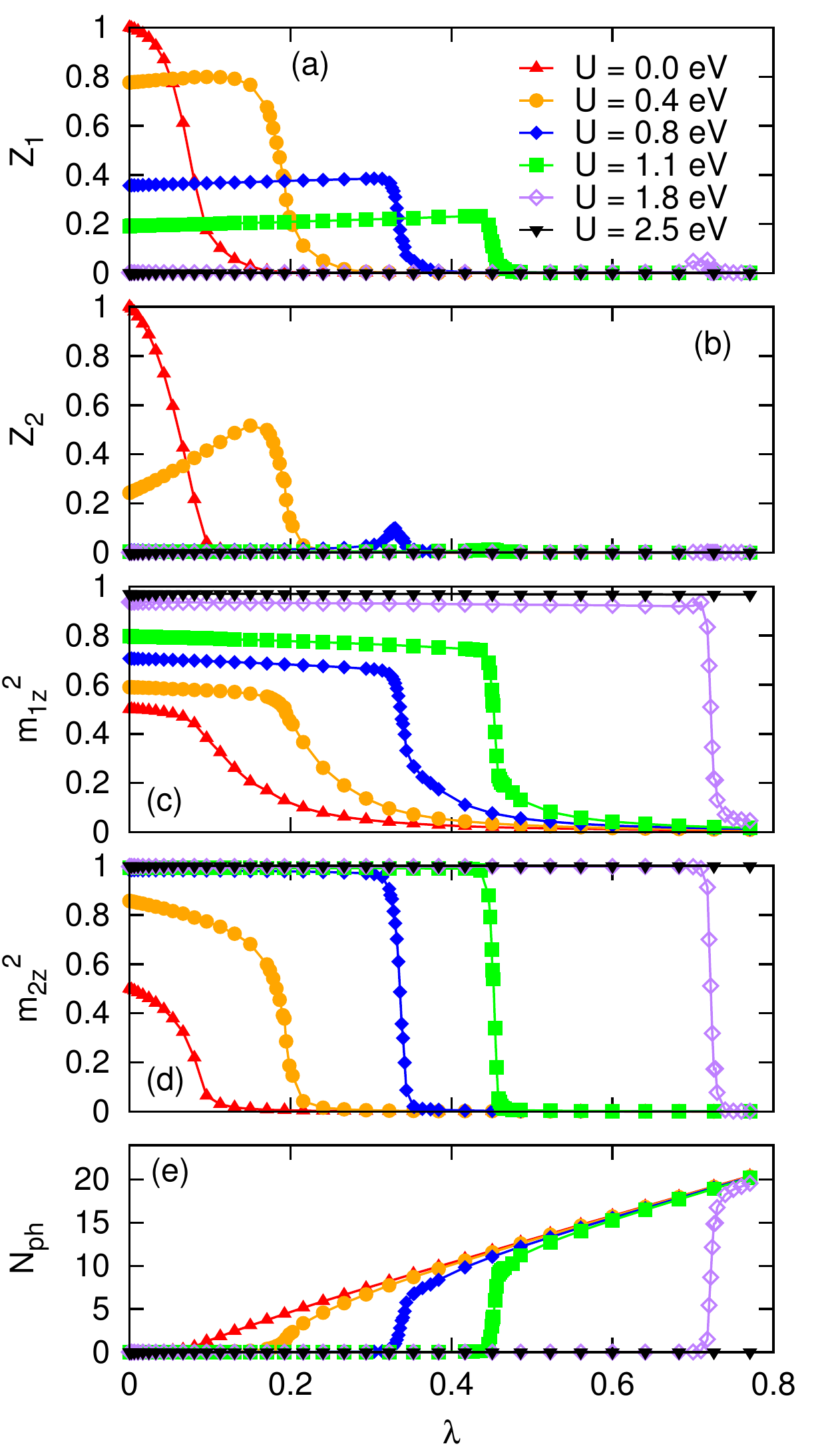}
\caption{\label{Fig:Z_and_S}{(color online)}  The quasiparticle weights (a)
$Z_{1}$  and (b) $Z_{2}$ as a function of the {\eph} interaction strength ($\lambda$) at different Hubbard U values. Mean
values of the local magnetic moments $m_{1z}^2$,
$m_{2z}^2$ and phonon numbers ($N_\mathrm{ph}$) are shown in (c), (d), and (e), respectively.}
\end{figure}

We studied the model using single-site DMFT \cite{Georges_mod}, with exact
diagonalization (ED) \cite{ED_review} as the impurity solver. DMFT maps the
full lattice model with local interactions onto an impurity model embedded in a
self-consistently determined host. In this case, the host is 
approximated using a set of  
$N_\mathrm{b}=4$ discrete energy levels (results for $N_\mathrm{b} = 6$ are
shown in the supplementary materials \cite{Supplement}, where we find good
convergence). We work in infinite dimensions (where DMFT is exact) by adopting
a Bethe lattice with a semi-circular density of states   
$\rho_\gamma(\epsilon)=\frac{8}
{\pi}\sqrt{\left(W_\gamma/2\right)^2-\epsilon^2}/W^2_{\gamma}$, where
$W_\gamma=4t_\gamma$ is the bandwidth. Throughout the paper, we set $W_1 = 5W_2
\equiv W = 2$ eV, fix the temperature at $T=\frac{1}{\beta}=0.01$ eV, unless
otherwise stated, and set the phonon energy to $\Omega = 0.15$ eV. (Results for
smaller values of $\Omega$ are qualitatively similar and can be found in 
Ref. \cite{Supplement}.) 
The bandwidth and Hund's coupling $J$ are chosen so that we can obtain a robust OSMP without the {\eph} coupling.
The dimensionless \eph~coupling constant is defined as
$\lambda=\frac{2g^2}{W\Omega}$.  The infinite phonon Hilbert space for the
impurity model is limited by only allowing up to $N_\mathrm{ph}$
phonons, where $N_\mathrm{ph}\sim 40$ is typical, depending on the parameters
used. We have checked that all of our results are well converged for increasing
values of $N_\mathrm{ph}$. 
 
{\em Results} --- The $\lambda$-$U$ phase diagram for the model is shown in
Fig. \ref{Fig:phasegu}.  Here, we plot the orbitally resolved Matsubara
quasiparticle weight $Z_{\gamma}=\left(1-\frac{\mathrm{Im}\Sigma({\rm i}\pi
T)}{\pi T}\right)^{-1}$ on a logarithmic scale. Five distinct phases can be
identified from the values of $Z_\gamma$, the local magnetic moment $m_{\gamma
z}^2=\langle \left( n_{\gamma \uparrow}-n_{\gamma \downarrow}\right)^2\rangle$,
and the average number of phonon quanta $N_\mathrm{ph} = \langle b^\dagger
b\rangle$ (all shown in Fig. \ref{Fig:Z_and_S}), and their  boundaries are
indicated by the white lines.  Three of these phases are similar to those found
in the single-band Hubbard-Holstein model.  The first phase is a metallic phase
(M) at small $(\lambda, U)$, where both $Z_1$ and $Z_2$ are large.  The second
is a Mott insulating (MI) phase, which appears at large $U$. It is identified
by a situation where $Z_\gamma = 0$, the magnetic moments are large $m_{1,z}^2
\approx m_{2,z}^2 \approx 1$, and $N_\mathrm{ph}$ is nearly zero. The third
phase is a CDW insulating phase where $Z_1=Z_2 = 0$, while $N_\mathrm{ph}$ is
large ($N_\mathrm{ph}\gg1$) and no local moments have formed (i.e. $m_{1,z}^2 \approx m_{2,z}^2 \approx 0$).  An examination of the wavefunctions reveals that the
CDW phase corresponds to a state where the impurity site is either fully
occupied or entirely empty with equal probability, consistent with a
checkerboard-type ordering common to the single-band model 
\cite{Nowadnick_PRL,Bauer_prb,Bauer_epl}.  This phase is likely to be a
$\left(\pi,\pi,...\right)$ CDW order (sometimes referred to as a strong
coupling bi-polaronic insulating phase in the single-band case). Alternatively,
this phase could also reflect phase separation, although delocalization effects
should favor the CDW. Further studies on extended clusters will be
needed to address this issue.

In addition to the ``standard" phases, we also observe two distinct phases with
orbital selective characteristics. The first is the widely studied OSMP, which
appears between the M and MI phases.  It resembles the same OSMP found in the
model without {\eph} interactions \cite{deMediciPRL2011}. Here, the orbital
with the narrower bandwidth becomes insulating with $Z_2 = 0$ and $m_{2z}^2
\approx 1$, while the orbital with the wider bandwidth remains itinerant with a
non-zero quasiparticle weight. Interestingly, we also observe a second region
of orbital selective behavior, located in a small portion of parameter space
between the M/OSMP phases and the fully insulating CDW phase, denoted as OSPI
in Fig. 1.  As with the OSMP, in this region, the narrow band becomes
insulating while the wide band remains itinerant with $Z_1 \neq 0$ and $Z_2
=0$. But unlike the OSMP, here we find tiny local moments on orbital 2 with
$m_{2 z}^2\le0.05$, and a large $N_\mathrm{ph}$ ($N_\mathrm{ph}>1$). The latter
indicates the presence of a sizable lattice distortion. The {\eph} interaction
drives the orbital-selective insulating properties in this case rather than the
Hubbard and Hund's interaction.  We label this state an orbital selective
Peierls insulator (OSPI), in analogy to the OSMP. 

For reference, Fig. \ref{Fig:Z_and_S} shows the evolution of the 
quantities used to identify the five regions of the phase diagram as a
function of $\lambda$ for different values of $U$. When $U\le0.4$ eV,
$m_{\gamma z}^2$ and $N_\mathrm{ph}$ vary smoothly near the phase transition,
while for $U>0.4$ eV, these quantities vary quickly in the transition region,
but are nevertheless continuous.  This behavior is consistent with a previous
DMFT study of the single band Hubbard-Holstein model \cite{Bauer_prb}, 
where a smooth transition occurs at weak coupling that becomes increasingly sharp as $U/W$ increases. 

\begin{figure}[t]
\center\includegraphics[width=0.75\columnwidth]{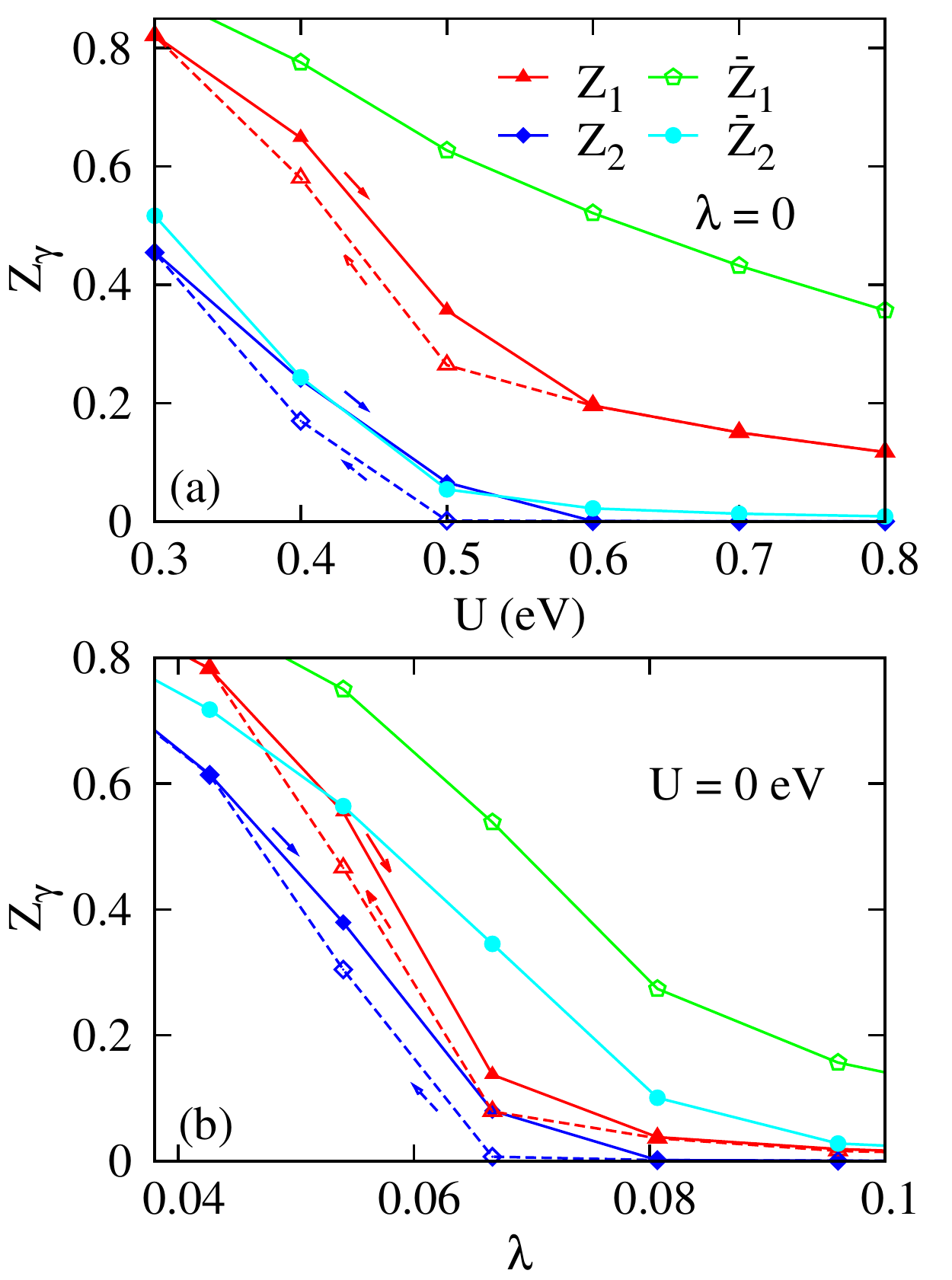}
\caption{\label{Fig:hys}{(color online)} (a) The quasiparticle weight $Z_{\gamma}$ as a function of $U$ at a fixed $\lambda=0$. (b) The quasiparticle weight $Z_{\gamma}$ as a function of $\lambda$ at a fixed $U=0$. $Z_{\gamma}$ are results at $T=0.002$ eV and $\bar{Z}_{\gamma}$ are results at $T=0.01$ eV. The solid lines and the dashed lines are results of increasing and decreasing $U$ or $\lambda$, respectively.}
\end{figure}
       
To study the analogy between the OSMP and the OSPI further, we examine the
classification of the phase transitions and their possible hysteresis behavior \cite{Liebsch}. Fig. \ref{Fig:hys}(a) and \ref{Fig:hys}(b) plot the evolution of $Z_{\gamma}$ at $T=0.002$ eV along the $(U,\lambda =0)$
and $(U = 0, \lambda)$ axes, respectively. Although there are two Mott
transitions in the two-orbital system, we observe a single hysteresis loop near
the OSMP boundary, which indicates a coexistence region, as discussed in Ref.
\cite{Liebsch}. The critical $U$ values for increasing and decreasing 
interaction strengths are $U_{c,1} = 0.6$ eV and $U_{c,2} = 0.5$ eV,
respectively. Similarly, along the ($U = 0,\lambda$) line we also find a single
coexistence region, consistent with DMFT studies for the single band Holstein
model \cite{Meyer_PRL, Jeon_PRB}. As with the Mott transition, the hysteresis
loop appears close to the first Peierls transition and the critical $\lambda$
values for increasing and decreasing interactions are $\lambda_{c,1} = 0.08$
and $\lambda_{c,2} = 0.066$, respectively. Thus, the OSMP and OSPI transitions
phenomena appear to be analogous to one another. The appearance of hysteresis
indicates a first order transition out of the metallic phase while the other
transitions are continuous. Finally, we note that the hysteresis behavior
disappears at $T=0.01$ eV, where we performed most of our calculations.

\begin{figure} 
\center\includegraphics[width=0.99\columnwidth]{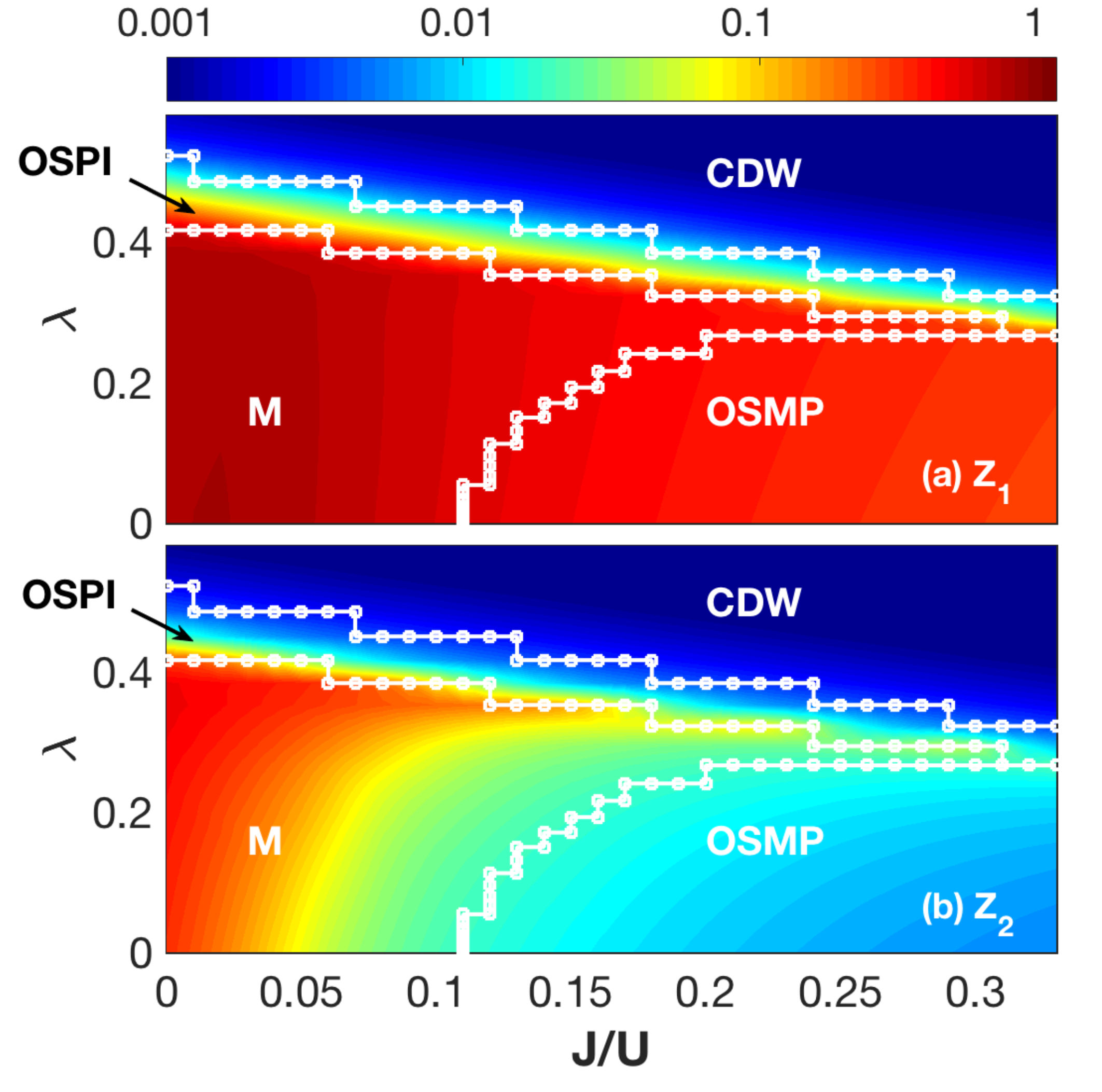}
\caption{\label{Fig:phasegJ}{(color online)} The phase diagram in the $\lambda$-$J/U$ plane at filling $n=2$. (a) and (b) plot quasiparticle weights $Z_{1}$ and $Z_{2}$, respectively. The labels used in this graph are the same as in Fig. \ref{Fig:phasegu}. The Coulomb interaction is fixed at $U=0.8$ eV and $U'=U-2J$. The white dots indicate points where the calculations were performed, and we plotted them to show phases boundaries. The color scale is plotted using a linear interpolation.}
\end{figure}

The Hund's coupling plays a major role in establishing the
boundaries of the OSMP \cite{Koga,Medici_PRL}. Therefore, we explored its role
in determining the CDW and OSPI phases observed here. Fig. \ref{Fig:phasegJ}
shows the phase
diagram in the  $\lambda$ - $J/U$ plane for a fixed $U=0.8$ eV. For
$\lambda<0.3$, the metallic phase survives to larger values of $J/U$ as $\lambda$ increases. This result is
consistent with the notion that the {\eph} interaction mediates an effective attractive interaction that competes with the onsite Hubbard interactions. 
When $0.3<\lambda<0.4$, the OSMP disappears and is replaced by the OSPI and
CDW phases and the critical $\lambda$ value for both phases is decreased as
$J/U$ increases. For larger $\lambda$, the CDW phase persists for all 
$J/U$ values. Thus, the Hund's coupling not only favors the OSMP transition but also has a stabilizing effect for the lattice-driven phases. 

In the single band Hubbard-Holstein model, the action of the repulsive Hubbard interaction and the effective attractive interaction mediated by the phonons gives rise to the competition between CDW and MI phases. Here, in the multi-orbital case, the stabilization of the CDW phase with increasing $J/U$ is due to the reduction of the interorbital Hubbard interaction, imposed by the condition that 
$U^\prime=U-2J$. In short, increasing $J$ reduces $U^\prime$ and
therefore also reduces the total potential energy cost for a double occupation
of a given site. The cost for creating a charge ordered phase, where each
site alternates between fully occupied and empty, is therefore lowered. This
interpretation can be confirmed explicitly by holding $U$ and $U^\prime$ fixed
while varying $J$. The corresponding phase diagram does not show the same
stabilization of the CDW phase with increasing $J$ \cite{Supplement}.

We have studied the interplay between the
{\ee} and {\eph} interactions in a degenerate two-orbital Hubbard-Holstein
model. A Competition between the onsite {\ee} and {\eph} interactions leads
to many competing phases including the OSMP and OSPI at small couplings and the
MI and CDW at large couplings.  We also find that the Hund's coupling $J$ has
nontrivial effects on the phases driven by the {\eph} interactions. Importantly,
our results demonstrate that weak to intermediate {\eph} interaction strengths can
have a significant impact on the phase diagram of this model. As such, one
cannot rule out an important role for phonons {\em a priori} in multi-orbital
systems, where multiple electronic interactions are already competing with one
another.  

We close with a short note and some speculation. Ref. \cite{Streltsov_PRB} has also used the term OSPI in the context of a two-orbital dimer model, where superexchange is stronger between a particular subset of orbitals, creating a preferential dimerization.  An entirely different mechanism drives our OSPI, where we start from a metallic state and obtain the OSPI through the {\eph} interaction. To the best of our the knowledge, this is the first time that theoretical calculations have produced such a mechanism. As with OSMP, the OSPI, in this case, is induced by the different bandwidths for the two orbitals. Finally, although the OSPI discovered here was derived from a Holstein coupling, we believe bond-stretching phonons that modulate interatomic hopping integrals could induce a similar phenomenon. In such cases, these interactions could have a significant impact on nematic phases observed in some FeSCs \cite{Sun_nature, Chu_science}.  

{\em Acknowledgments} --- 
This work was supported by the University of Tennessee's Science Alliance Joint Directed Research Development (JDRD) program, a collaboration with Oak Ridge National Laboratory. S. J. also acknowledges support from the German Research Foundation (DFG) under SFB 1143 and hospitality of the IFW-Dresden. E. K. is supported by the National Science Foundation Grant No. DMR-1609560. CPU time was provided by resources supported by the University of Tennessee and Oak Ridge National Laboratory Joint Institute for Computational Sciences (http://www.jics.utk.edu).

\end{document}